\documentclass[10pt,twocolumn,a4paper,fleqn]{article}
\usepackage[a4paper,left=15mm,right=15mm,top=30mm,bottom=25mm,columnsep=10mm]{geometry}
\usepackage{graphicx}
\usepackage{times}
\usepackage{titlesec}
\usepackage{indentfirst}
\usepackage[fleqn]{amsmath}
\usepackage{fancyhdr}
\usepackage{cite}

\renewcommand{\normalsize}{\fontsize{10pt}{11.9pt}\selectfont} 
\titleformat{\section}
  {\normalfont\large\bfseries\MakeUppercase}{\thesection.}{0.5em}{}
\titlespacing*{\section}{0pt}{2.5ex plus 0.5ex minus .2ex}{1.5ex}

\titleformat{\subsection}
  {\normalfont\normalsize\bfseries}{\thesubsection.}{0.5em}{}
\titlespacing*{\subsection}{0pt}{1.5ex}{0.5ex}

\titleformat{\subsubsection}
  {\normalfont\normalsize}{\thesubsubsection.}{0.5em}{}
\titlespacing*{\subsubsection}{0pt}{2.0ex plus .5ex minus .2ex}{0.1ex}

\usepackage[small,bf]{caption}
\DeclareCaptionLabelSeparator{periodspace}{.~ }
\makeatletter
\newcommand{\affils}[1]{\def\@affils{#1}}
\renewcommand{\abstract}[1]{\def\@abstract{#1}}
\newcommand{\keywords}[1]{\def\@keywords{#1}}

\renewcommand{\@maketitle}{%
  \newpage
  \null
  \begin{center}
    {\fontsize{15pt}{15pt}\selectfont \bfseries \@title \par}
    \vskip 1.0em
    {\large \@author \par}
    \vskip 0.5em
    {\normalsize \@affils \par}
  \end{center}
  \vskip 1em
  {\normalsize \noindent \textbf{Abstract:} \@abstract \par}
  \vskip 1em
  {\normalsize \noindent \textbf{Keywords:} \@keywords \par}
  \vskip 1.4em
}
\makeatother

\makeatletter
\renewenvironment{thebibliography}[1]{
  \section*{\refname}
  \normalsize
  \list{[\arabic{enumi}]}{
    \settowidth\labelwidth{[#1]}
    \leftmargin\labelwidth
    \advance\leftmargin\labelsep
    \setlength{\itemsep}{0pt}
    \setlength{\parsep}{0pt}
    \setlength{\topsep}{0pt}
    \setlength{\partopsep}{0pt}
    \usecounter{enumi}
    
  }
  \def\newblock{\hskip .11em plus .33em minus .07em}
  \sloppy\clubpenalty4000\widowpenalty4000
  \sfcode`\.=1000\relax
}{
  \endlist
}
\makeatother

\usepackage{bm}
\usepackage{subcaption}
\usepackage{amsmath}
\usepackage{url}
\newcommand{\mypara}[1]{%
  \vspace{1mm}%
  \noindent\textbf{#1}\quad
}

\fancypagestyle{firstpage}{
  \fancyhf{}
  
  \fancyfoot[C]{\footnotesize \sffamily Author's version. In Proc. of the Joint Symposium of AROB 31st and ISBC 11th (AROB-ISBC 2026), pp.~1620--1624, 2026.}
}

\title{Modeling and Analysis of Fish Interaction Networks\\under Projected Visual Stimuli}

\author{Hiroaki Kawashima${}^{1}$, Raj Rajeshwar Malinda${}^{1}$, and Saeko Takizawa${}^{1}$}

\affils{${}^{1}$Graduate School of Information Science, University of Hyogo, Hyogo, Japan\\
(E-mail: kawashima@gsis.u-hyogo.ac.jp, rrmalinda@gsis.u-hyogo.ac.jp, s.takizawa.student@gmail.com)\\
}
\abstract{%
This paper addresses the estimation of a dynamic interaction network, a network of influence among individuals, under projected visual stimuli to quantify the influences of inter-individual interactions and external stimuli on collective behavior.
Building upon our previously proposed network estimation model, which assumes a Boids-type model and employs a sparse regression framework to infer inter-individual influence networks from trajectory data, we extend the formulation by introducing a stimulus term. This enables the model to capture how individuals react to and propagate externally projected visual stimuli within the group. The resulting framework allows simultaneous estimation of inter-individual and stimulus-related interaction strengths. We also introduce entropy-based indices to capture the possible biases of individuals' influence. Our experiments with fish schools under projector-based visual stimuli demonstrate the effectiveness of the proposed indices in quantifying schooling behavior and identifying influential individuals within the group, serving as the basis for real-time, interpretable metrics of collective dynamics.
}


\begin{document}

\maketitle
\thispagestyle{firstpage}

\section{Introduction}
\label{sec:introduction}

Identifying influential individuals within a swarm and understanding how external interventions modulate collective behavior is crucial for developing ``swarm–machine interaction systems.'' Such systems enable artificial agents to interact adaptively with biological collectives, such as fish schools. While recent advances in tracking and modeling have enabled high-resolution observation of collective motion, a critical gap remains in the real-time, quantitative identification of dynamic leadership transitions and the evaluation of stimulus-induced effects. 
Addressing this challenge requires interpretable, model-based indices that characterize both group- and individual-level dynamics during behavioral transitions. This study proposes such indices derived from a minimal interaction model that explicitly incorporates inter-individual influence and external stimulus components.

To estimate interaction networks, many researchers infer connectivity from instantaneous spatial features, such as relative distance or field-of-view constraints~\cite{rosenthalRevealingHiddenNetworks2015}. Although these geometric criteria are intuitive and widely used, they may not capture the underlying causal structure of a swarm. For predicting future collective dynamics, it is more desirable to quantify the direct influence between individuals, capturing the causal impact of one agent on another, rather than relying on static spatial configurations.

Predictive formulations define interactions based on how much one individual's behavior facilitates the prediction of another's future state. The core difficulty stems from the fact that these underlying influence strengths are latent and must be estimated from time-series data. While representative techniques, such as vector autoregressive models~\cite{bolstadCausalNetworkInference2011} and regularized neural networks~\cite{marcinkevicsInterpretableModelsGranger2021,fujiiLearningInteractionRules2021}, effectively handle complex nonlinear or high-dimensional dependencies, they often lack the interpretability essential for understanding the fundamental mechanisms of collective animal behavior.

In response to these challenges, we previously introduced an interpretable, data-driven framework to estimate dynamic interaction networks from trajectory data~\cite{Kawashima23}. By adopting a simple linear formulation and sparse regression, this method maintains simplicity while effectively capturing time-varying inter-individual influences. This approach is designed to uncover how individuals integrate environmental information, revealing which specific features of their neighbors most strongly influence their behavior.

The inferred network offers key insights for analyzing information propagation and predicting collective behavior. It also provides a framework for designing interventions to effectively guide the entire group~\cite{rahmaniControllabilityMultiAgentSystems2009,kawashimaManipulabilityLeaderFollower2014,egerstedtInteractingNetworksMobile2014a}.

Building on the framework in \cite{Kawashima23}, this study introduces an extended model that explicitly accounts for the direction of visual stimuli at each individual's position. The objective is to disentangle internal group coordination from external environmental factors, thereby quantifying their relative contributions to the collective motion. By incorporating this additional term, we can more effectively characterize how external visual cues modulate the interaction dynamics within the group.

\section{Proposed model}
\label{sec:proposed_model}

\subsection{Interaction model under external stimuli}
\label{sec:model_with_stimuli}

We first briefly review the base interaction model proposed in~\cite{Kawashima23}.
We assume that the total number of individuals is fixed at $N$. Let $\bm{p}_i(t) \in R^d$ be the position (centroid) of individual $i$ ($i=1, \ldots, N$), where $d=2$ (planar) in this study.
Letting $\bm{v}_i(t) = \dot{\bm{p}}_i(t)$ denote its velocity, we model the dynamics of individual $i$ as follows:
\begin{equation}
    \bm{v}_i(t+\tau)
    = \sum_{j \neq i} w_{ij}(t) \hat{\bm{p}}_{ij}(t) 
    + \bm{d}_i(t) + \bm{\epsilon}_i(t+\tau).
    \label{eq:continuous_time_model}
\end{equation}
Here, $\hat{\bm{p}}_{ij}(t)$ is the normalized direction vector from individual $i$ to $j$, defined as $\bm{p}_{ij}(t) = \bm{p}_j(t) - \bm{p}_i(t)$; throughout this paper, the notation $\hat{\bm{x}}$ denotes the normalized vector $\bm{x} / \|\bm{x}\|$.
The weights $w_{ij}(t) \geq 0$ quantify the strength of the influence of individual $j$ on $i$, representing the attractive component of their coordination; $w_{ij}(t) = 0$ signifies the absence of influence.
The term $\bm{d}_i(t) = w_{ii}(t) \hat{\bm{d}}_i(t)$ represents the autonomous component, which accounts for $i$'s preferred direction, with $w_{ii}(t) \geq 0$ signifying its magnitude. The remaining term $\bm{\epsilon}_i(t+\tau)$ is noise, and the parameter $\tau > 0$ represents a short time lag to account for temporal delays in response.
While the model can be extended to include an alignment term as in the original Boids model~\cite{Reynolds1987}, we omit it here for simplicity.

To incorporate visual stimulus effects, we extend Eq.~\eqref{eq:continuous_time_model} by adding the term $w_{i}^{\mathrm{stim}}(t) \hat{\bm{s}}_i(t)$, where $w_{i}^{\mathrm{stim}}(t) \ge 0$ is a weight representing the strength of the stimulus effect on individual $i$. The vector $\hat{\bm{s}}_i(t)$ denotes the normalized direction vector of the visual stimulus at individual $i$'s position at time $t$. 
Discretizing this extended model with an observation cycle $T_o$ yields:
\begin{equation}
    \bm{v}_{i,t+\tau_d}
    = \sum_{j\ne i} w_{ij,k}\,\hat{\bm{p}}_{ij,t}
      + w_{ii,k}\hat{\bm{d}}_{i,k}
      + w_{i,k}^{\mathrm{stim}}\,\hat{\bm{s}}_{i,t}
      + \bm{\epsilon}_{i,t+\tau_d},
      \label{eq:discrete_with_stimulus_model}
\end{equation}
where $\tau = T_o \tau_d$. By an abuse of notation, we re-index the time points $0, T_o, 2T_o, \ldots$ as $t = 0, 1, 2, \ldots$.
The index $k$ denotes the time window for parameter estimation; in this study, we set $k = t$ (i.e., a sliding window with a step size of 1). The parameters to be estimated for each individual $i$ within window $k$ are the weights $(\{w_{ij,k}\}_{j}, w_{i,k}^{\mathrm{stim}})$ and the preferred direction $\hat{\bm{d}}_{i,k}$.

The estimation procedure follows the sparse regression framework described in \cite{Kawashima23}, with the modification that the design matrix now explicitly includes the stimulus-direction term.

\subsection{Schooling indices under visual stimuli}

The parameters estimated from Eq.~\eqref{eq:discrete_with_stimulus_model} allow for the derivation of various indices to quantify schooling behavior under visual stimuli. In this section, we define several key metrics designed to facilitate a detailed analysis of the underlying factors driving collective dynamics.

\subsubsection{Group-level indices}

Based on the parameters estimated in Eq.~\eqref{eq:discrete_with_stimulus_model}, we define two group-level indices: \textit{coordination strength}, which quantifies the overall degree of inter-individual coordination, and \textit{stimulus responsiveness}, which represents the collective sensitivity to external visual stimuli. 
Specifically, we define these indices as the sums of the estimated coefficients across all individuals:
\begin{align}
 \label{eq:coordination_strength}
& \text{Coordination strength:} & S_{\mathrm{att}} &= \sum_i \sum_{j\neq i} w_{ij}, \\
 \label{eq:stimulus_responsiveness}
& \text{Stimulus responsiveness:} & S_{\mathrm{stim}} &= \sum_i w_{i}^{\mathrm{stim}},
\end{align}
where the time index $k$ is omitted for clarity.

The coordination strength $S_{\mathrm{att}}$ reflects the total magnitude of inter-individual interactions within the group; a higher value corresponds to a stronger tendency toward coordinated schooling. 
Similarly, $S_{\mathrm{stim}}$ represents the group's overall sensitivity to external visual stimuli, derived from the total weight of the stimulus terms across all individuals.

\subsubsection{Individual influence within a group}

While the group-level indices focus on the ``responses'' of individuals to their neighbors and the environment, we also evaluate how each individual influences the rest of the group. 
To quantify this, we define the \textit{individual influence} $I_i$ that individual $i$ exerts on others as the sum of the weights of the edges originating from $i$:
\begin{equation}
    I_{i} = \sum_{j \ne i} w_{ji},
    \label{eq:individual_influence}
\end{equation}
where the time index $k$ is omitted here as well.

By monitoring the dynamic evolution of these influence scores within the estimated network, we can identify key individuals who play dominant roles in the group's behavior at any given time. This approach also allows us to track how influence propagates through the collective. 
Furthermore, because the proposed framework is computationally efficient, these indices can be calculated in real time. These characteristics make our method particularly suitable for applications requiring dynamic interventions, such as guiding a collective by targeting and stimulating its most influential members.

\subsubsection{Individual variation of influence}

We also investigate whether the group contains specific ``core'' individuals that exert a disproportionately strong influence on others. To characterize the diverse roles individuals play within the school, we introduce an entropy-based index to analyze the variation in influence among individuals. Specifically, we calculate the normalized entropy of the time-averaged relative influence to quantify the degree to which influence is centralized or uniformly distributed within the group.

For each time step $k$, we normalize the individual influence $I_{i,k}$ as $r_{i,k} = I_{i,k} / \sum_j I_{j,k}$, ensuring $\sum_i r_{i,k} = 1$. We then compute the time-averaged relative influence for each individual as
$
    \bar{r}_i = \langle r_{i,k} \rangle_k,
$
which satisfies $\sum_i \bar{r}_i = 1$ and represents the average proportion of influence exerted by individual $i$ over the observation period. %
\begin{equation}
    H_{\mathrm{influ}} = - \sum_{i=1}^{N} \bar{r}_i \log_{N} \bar{r}_i,
    \label{eq:entropy_influence}
\end{equation}
where the base $N$ for the logarithm ensures that the index is normalized to the range $[0, 1]$. A lower value of $H_{\mathrm{influ}}$ indicates that influence is concentrated in a few individuals, whereas $H_{\mathrm{influ}} = 1$ corresponds to a perfectly uniform distribution of influence. 

Similarly, we define $H_{\mathrm{stim}}$ to quantify the variation in stimulus responsiveness among individuals by replacing $I_{i,k}$ with the stimulus weight $w_{i,k}^{\mathrm{stim}}$ in Eq.~\eqref{eq:discrete_with_stimulus_model}.

\vspace{-1mm}
\section{Experiments}

\subsection{Speed-controlled visual stimuli}
\label{sec:stimuli_experiment}

To validate the proposed model and indices, we conducted projector-based behavioral experiments using schools of rummy-nose tetras. For each trial, five fish were placed in a shallow, square tank (40~cm $\times$ 40~cm, 5~cm deep). 
Following an acclimation period, clockwise-rotating stimuli were presented onto the bottom of the tank as visual stimuli at three distinct constant angular velocities (S1--S3).  Fish movements were recorded using an overhead color camera at 60~fps. These conditions were designed to induce rotational schooling behaviors, providing a controlled context for analyzing transitions between spontaneous (free) schooling and stimulus-induced schooling through the estimated interaction networks.

Individual trajectories were extracted using the YOLO11 multi-object tracking framework~\cite{yolo11_ultralytics} followed by manual correction to ensure accuracy. Experiments were conducted over three days with three trials (cycles) per day, covering four experimental conditions: control (C), slow rotation (S1: $0.286^{\circ}/\mathrm{frame}$), medium rotation (S2: $0.572^{\circ}/\mathrm{frame}$), and fast rotation (S3: $1.144^{\circ}/\mathrm{frame}$). Each condition lasted $40\,\mathrm{s}$, and the middle $35\,\mathrm{s}$ of each recording were used for analysis. The order of conditions was counterbalanced across trials using a Latin square design. Further details regarding the experimental setup and procedure are available in~\cite{RajBioRxiv2025}.

\mypara{Estimated individual coefficients}

To provide a foundation for our group-level analysis, we first examine the time-averaged coefficients for each individual under the various stimulus conditions. This allows us to understand how individual behavior is driven by inter-individual attractive coordination, external rotational stimulation, and intrinsic autonomous components. For the stimulus direction in Eq.~\eqref{eq:discrete_with_stimulus_model}, we set $\hat{\bm{s}}_{i,t} = \hat{\bm{r}}_{i,t}$, where $\hat{\bm{r}}_{i,t}$ denotes the tangential rotational direction of the dot pattern at individual $i$'s position at time $t$. 

\begin{figure}[t]
    \centering
    \begin{subfigure}[t]{0.48\linewidth}
        \includegraphics[width=\linewidth]{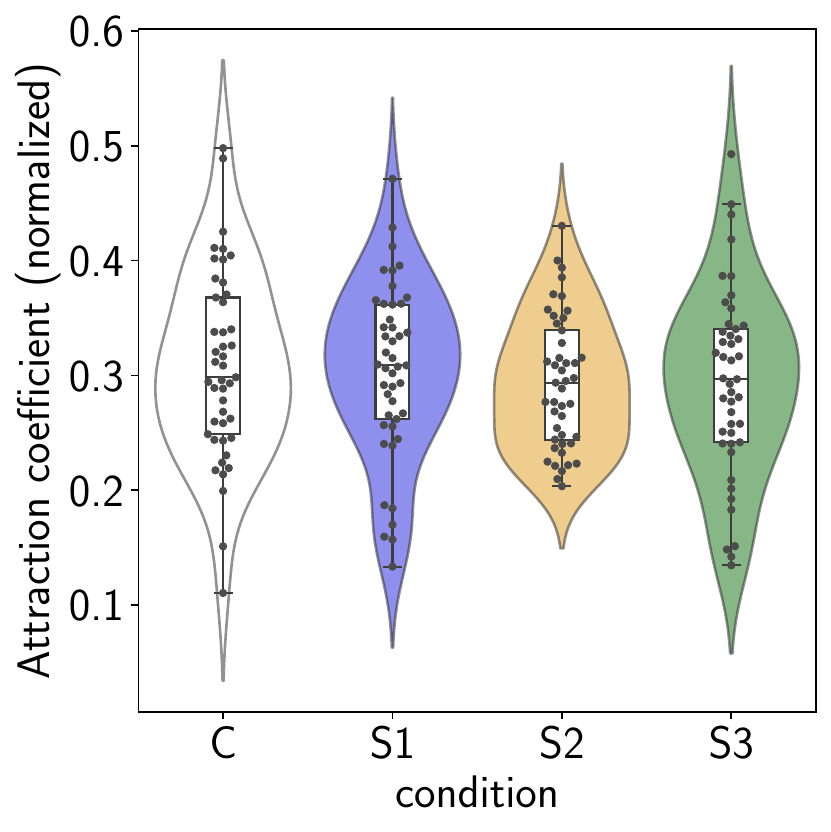}
        \caption{Coordination coefficient}
        \label{fig:indiv_coordinated_coef}
    \end{subfigure}
    \begin{subfigure}[t]{0.48\linewidth}
        \includegraphics[width=\linewidth]{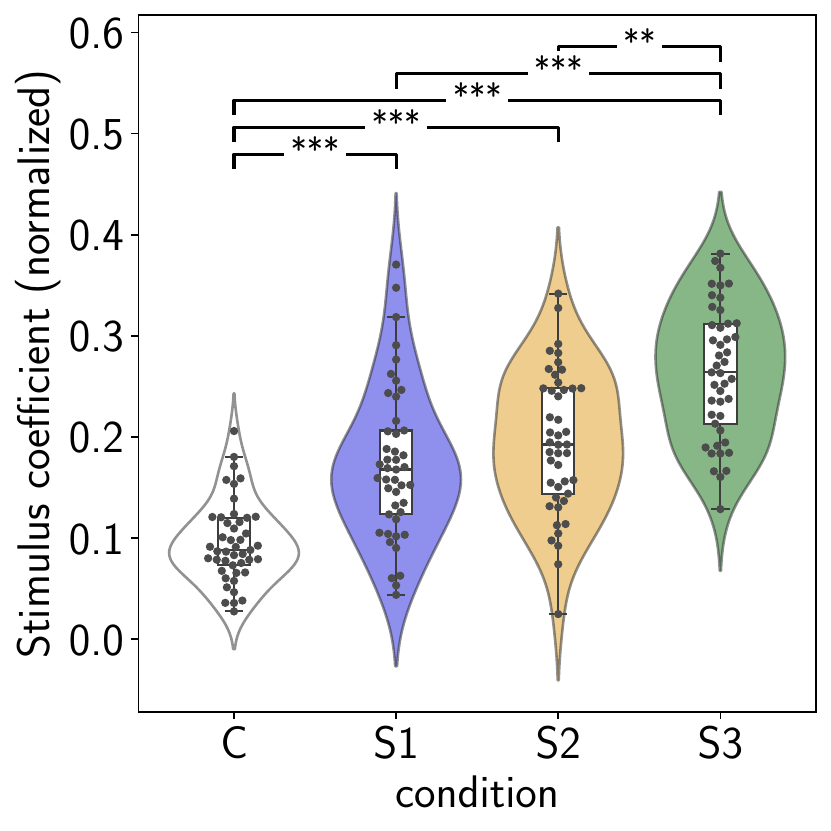}
        \caption{Stimulus coefficient}
        \label{fig:indiv_stimulation_coef}
    \end{subfigure}
    \begin{subfigure}[t]{0.48\linewidth}
        \includegraphics[width=\linewidth]{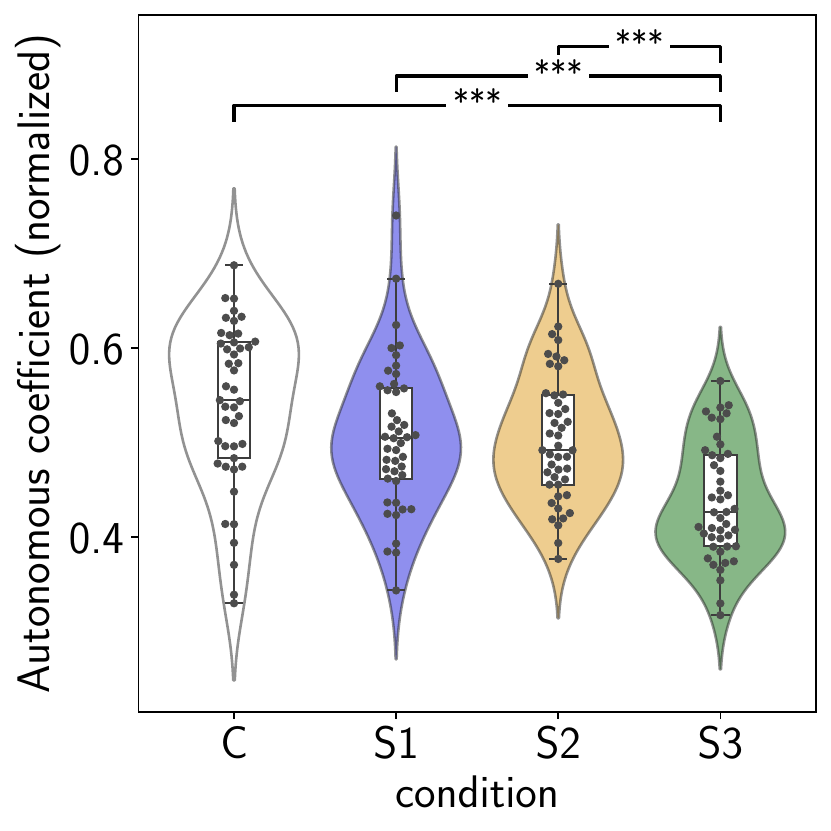}
        \caption{Autonomous coefficient}
        \label{fig:indiv_autonomous_coef}
    \end{subfigure}
    \caption{Distributions of time-averaged normalized individual coefficients for (a) coordination, (b) stimulus, and (c) autonomous components under conditions C and S1–S3. Statistical significance was assessed using the Kruskal-Wallis test, followed by Dunn's post-hoc test with Bonferroni correction (*$p < 0.05$, **$p < 0.01$, ***$p < 0.001$).}
    \label{fig:individual_coefficients}
\end{figure}

Figure~\ref{fig:individual_coefficients} presents the distributions of the time-averaged normalized coefficients for the coordination (attraction) component $\sum_{j\neq i} w_{ij}$, the stimulus component $w_i^{\mathrm{stim}}$, and the autonomous component $w_{ii}$ across conditions C, S1, S2, and S3. Because the raw coefficient values are scale-dependent on the velocity magnitude, we employed normalized values such that the sum of these three components for each individual at each time point equals 1. This normalization facilitates a direct comparison of the relative influence of each factor on individual behavior. To establish a baseline, the stimulus direction term $\hat{\bm{s}}_{i,t} = \hat{\bm{r}}_{i,t}$ was retained even in the no-stimulus control condition (C), allowing us to quantify the spontaneous tendency of the fish to rotate in the absence of external cues.

As shown in Fig.~\ref{fig:individual_coefficients}, the coordination terms remain relatively consistent across all experimental conditions. In contrast, behavioral shifts are prominent in the other components. Under the high-speed stimulus (S3), all individuals exhibit a marked increase in stimulus contribution. Conversely, in the control condition (C), the contribution of the autonomous component is notably larger. These results demonstrate that comparing time-averaged coefficients effectively captures the shift in behavioral drivers as the intensity of the external stimulus increases.

\subsection{Group-level schooling indices}

The group-level indices derived from our model provide quantitative insights into the factors that drive behavioral changes in individuals. To illustrate this qualitatively, we analyze a representative trajectory under the fast-rotation stimulus condition (S3).

\begin{figure}[t]
    \centering
    \includegraphics[width=\linewidth]{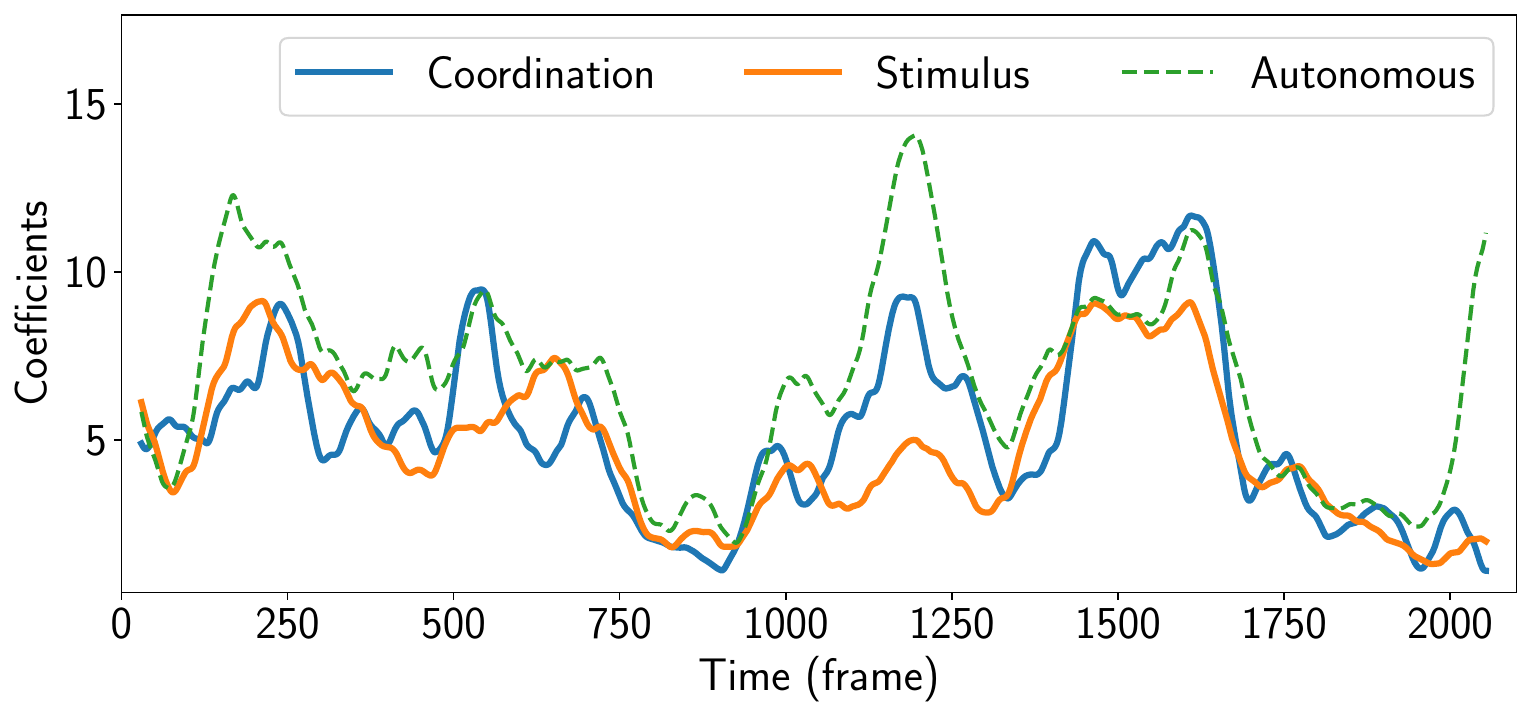}
    \caption{Time-series of model-derived indices under stimulus condition S3: coordination strength $S_{\mathrm{att}}$ (solid blue), stimulus responsiveness $S_{\mathrm{stim}}$ (solid orange), and the sum of autonomous components $\sum_i w_{ii}$ (dashed line).}
    \label{fig:schooling_coefsums_stim}
\end{figure}

Figure~\ref{fig:schooling_coefsums_stim} displays an example of the temporal evolution of $S_{\mathrm{att}}$ and $S_{\mathrm{stim}}$ (defined in Eqs.~\eqref{eq:coordination_strength} and \eqref{eq:stimulus_responsiveness}, respectively) under stimulus condition S3, along with the total autonomous contribution $\sum_i w_{ii}$ for reference. We observe that $S_{\mathrm{att}}$ remains relatively high during the initial period (frames 0--750), the middle period (frames 1100--1300), and around frames 1400--1700. Notably, while $S_{\mathrm{stim}}$ remains large during most of these intervals, it takes relatively lower values during frames 1100--1300.

This discrepancy indicates that during the 1100--1300 frame interval, the collective motion is dominated by internal group dynamics rather than the external stimulus. This interpretation is further supported by the relatively high values of the autonomous component around frame 1200.
A similar tendency is observed around frame 550 as well, while the discrepancy is less pronounced.
In such periods, subgroups of individuals often exhibit coordinated motion in directions, deviating from the rotational stimulus, which is clearly captured by the decoupling of our model-derived indices.

\subsection{Influence analysis}

To demonstrate the dynamic shifts in individual roles depending on the behavioral context and stimulus conditions, we analyze the temporal variation of the influence indices.

\begin{figure}[t]
    \centering
    \begin{subfigure}[t]{\linewidth}
        \includegraphics[width=\linewidth]{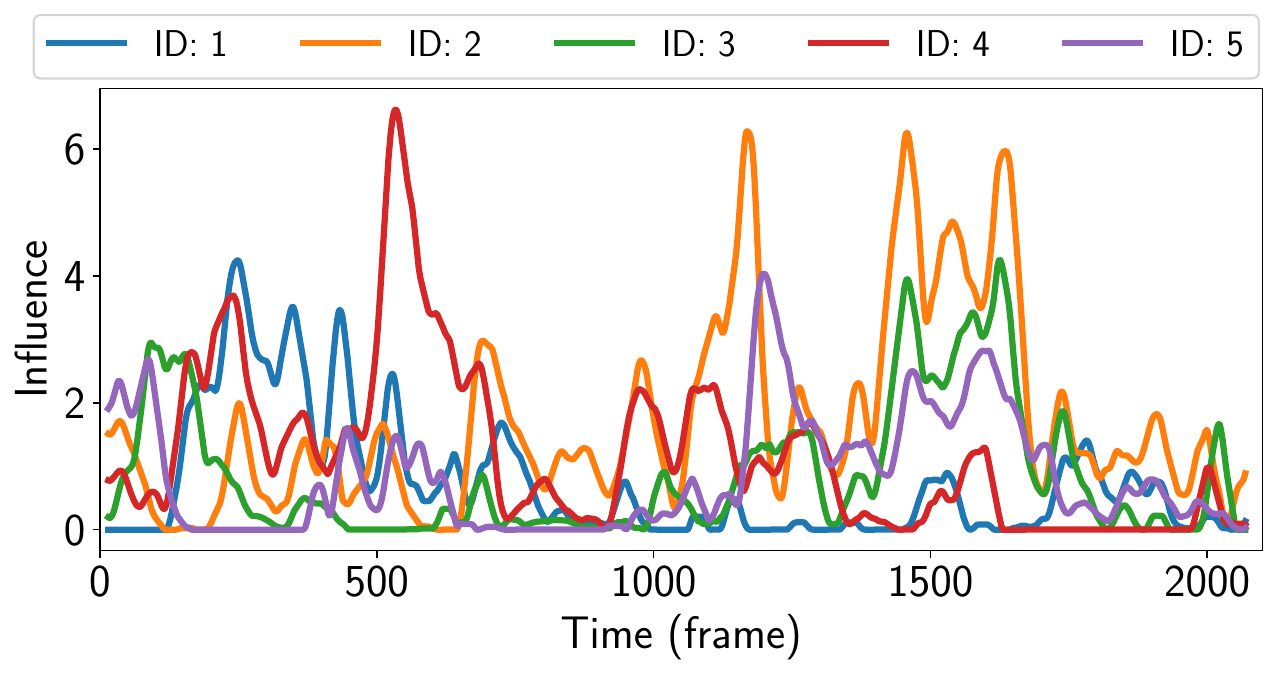}
        \caption{Temporal variation of individual influence.}
        \label{fig:fish_coef_influence}
    \end{subfigure}
    \begin{subfigure}[t]{\linewidth}
        \centering
        \includegraphics[width=0.48\linewidth]{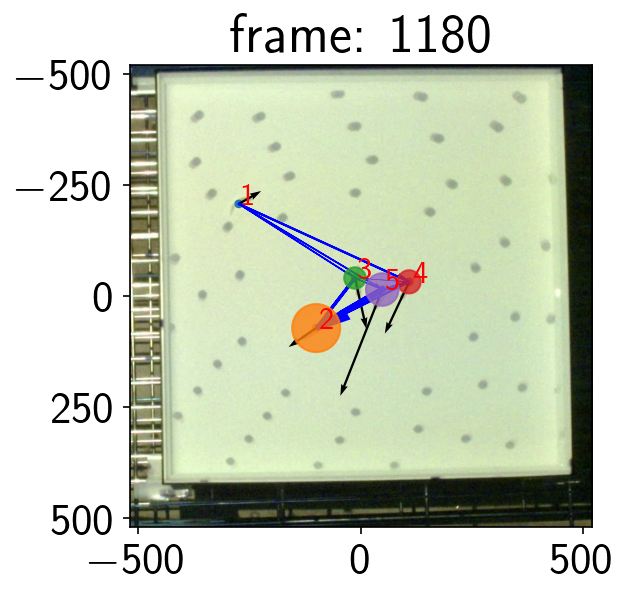}    
        \includegraphics[width=0.48\linewidth]{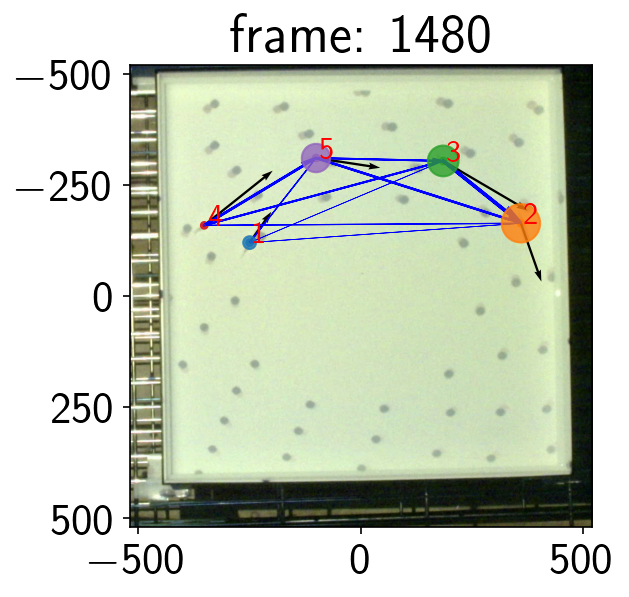}        
        \caption{Estimated interaction networks at frame 1180 (left) and 1480 (right).}
        \label{fig:estimated_networks}
    \end{subfigure}
    \caption{Individual influence analysis for condition S3. (a) Temporal variation of $I_i$. (b) Network snapshots where node size is proportional to $I_i$ and edge thickness reflects weight $w_{ij}$ (thresholded for clarity).}
    \label{fig:fish_influence_networks}
\end{figure}

Figure~\ref{fig:fish_coef_influence} shows the temporal variation of each individual's influence (defined in Eq.~\eqref{eq:individual_influence}), calculated from the same trajectory data as in Fig.~\ref{fig:schooling_coefsums_stim}. Around frames 1100--1200 and frames 1400--1700, ID~2 exerts a notably strong influence on the others. This is followed shortly by a peak for ID~5 around frame 1200. During these intervals, these specific individuals move in directions that deviate from the stimulus rotation; in particular, ID~2 leads the group in a ``shortcut'' behavior (Fig.~\ref{fig:estimated_networks}, left).
Later, around frame 1400, many fish begin to follow the external stimulus. Even in this different context, ID~2 again exerts a strong influence, effectively initiating the schooling motion along the rotational direction of the stimulus (Fig.~\ref{fig:estimated_networks}, right).

\begin{figure}[t]
    \centering
    \includegraphics[width=\linewidth]{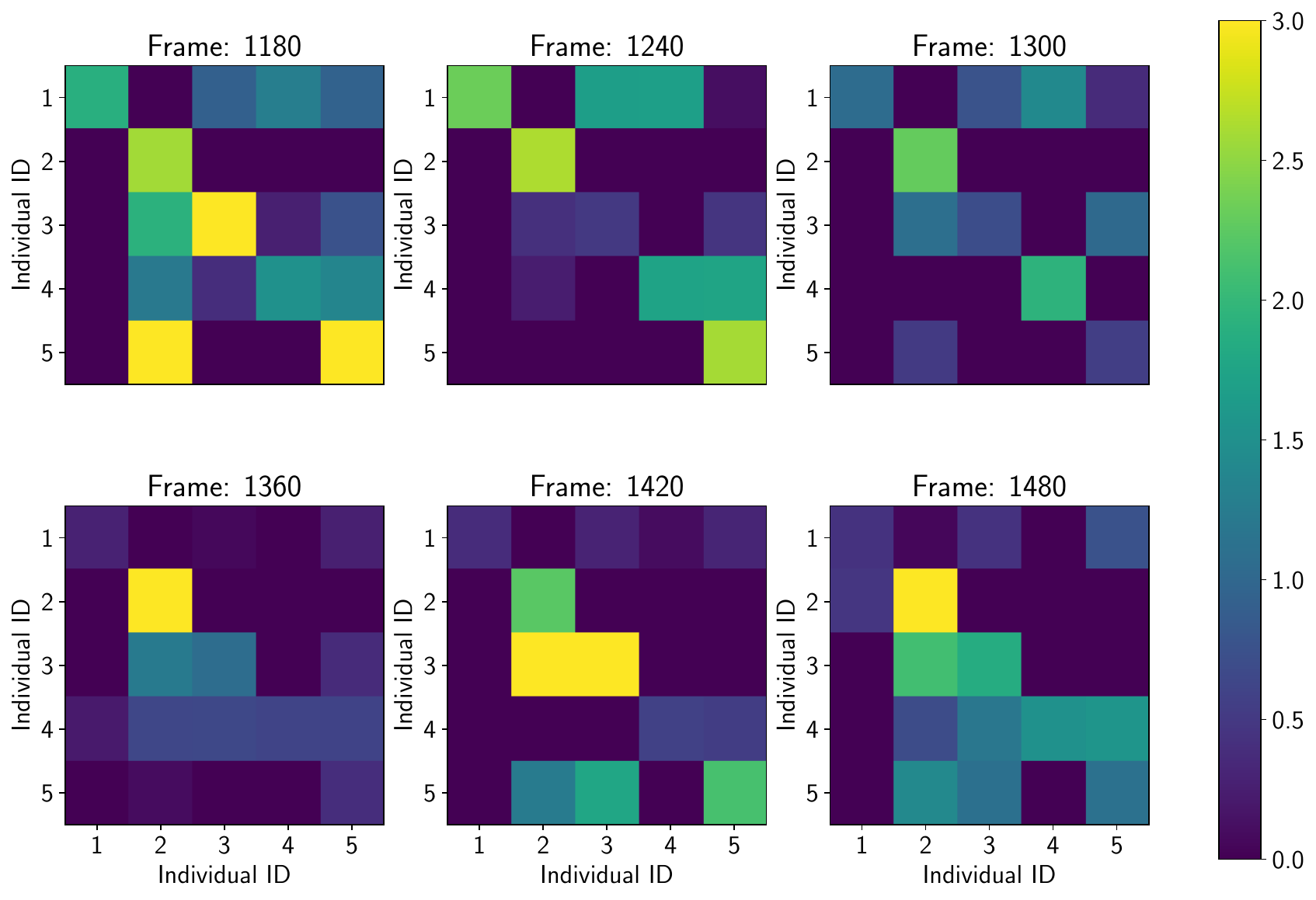}
    \caption{Estimated weight matrices $w_{ij}$ sampled every 60 frames (1~s) from frame 1180 to 1480. Diagonal and off-diagonal elements represent autonomous and inter-individual coordination (attraction) coefficients, respectively.}
    \label{fig:schooling_coef_matrices}
\end{figure}

A more detailed examination of inter-individual influence can be performed by revisiting the estimated weight matrices $w_{ij}$ visualized in Fig.~\ref{fig:schooling_coef_matrices}. In the snapshots for frames 1180, 1420, and 1480, the off-diagonal elements in the column for $j=2$ exhibit relatively large values, indicating which individuals are most influenced by ID~2.

\subsection{Leadership and behavioral homogeneity}

As observed in Fig.~\ref{fig:fish_coef_influence}, certain individuals (e.g., ID~2) transiently lead the group over short durations; however, such leadership is not fixed and appears to shift dynamically among individuals. This observation prompts the question of whether this shifting leadership is a general property of the group, characterized by a lack of systematic bias in influence, and how external stimulus strength modulates this tendency.

\begin{figure}[t]
    \centering
    \begin{subfigure}[t]{0.48\linewidth}
        \includegraphics[width=\linewidth]{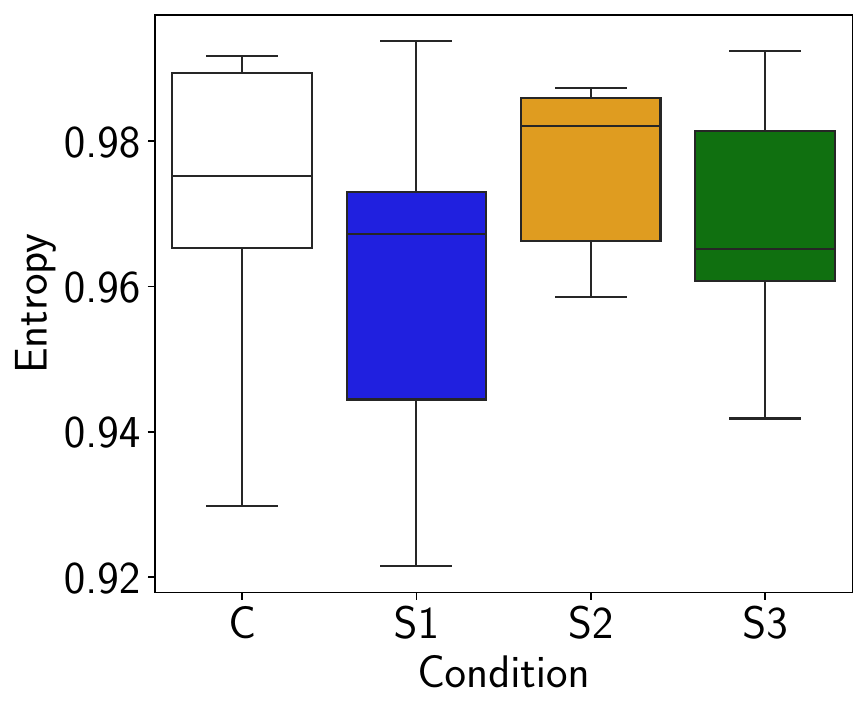}
        \caption{$H_{\mathrm{influ}}$}
        \label{fig:entropy_influence}
    \end{subfigure}
    \hfill
    \begin{subfigure}[t]{0.48\linewidth}
        \includegraphics[width=\linewidth]{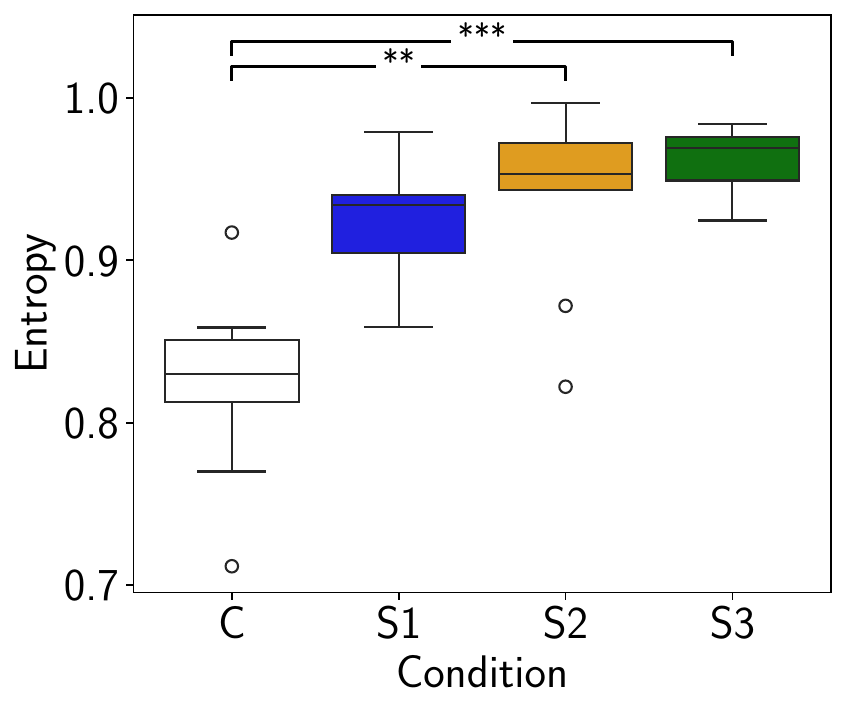}
        \caption{$H_{\mathrm{stim}}$}
        \label{fig:entropy_rotation}
    \end{subfigure}
    \caption{Normalized entropy of the time-averaged relative influence ($H_{\mathrm{influ}}$) and stimulus responsiveness ($H_{\mathrm{stim}}$) across the group. Statistical notation follows Fig.~\ref{fig:individual_coefficients}.}
    \label{fig:entropy_coefficients}
\end{figure}

Figure~\ref{fig:entropy_influence} shows that $H_{\mathrm{influ}}$ remains consistently near 1 across all conditions, with no significant differences. This indicates that over longer time scales, influence is distributed nearly uniformly across individuals, confirming the absence of a persistent leader. 
In contrast, the entropy of stimulus responsiveness, $H_{\mathrm{stim}}$ (Fig.~\ref{fig:entropy_rotation}), reveals marked individual variability in the control condition (C) but increases toward 1 under stronger stimulus conditions (S2, S3). 

This trend suggests that while individuals exhibit specific directional preferences or biases in the absence of a stimulus (C), stronger rotational stimuli induce a homogenization of behavioral responses across the group. Although not statistically significant, S1 exhibits greater inter-individual variability than S2 and S3, suggesting a higher degree of behavioral freedom at lower stimulus speeds. Conversely, the high-speed stimuli in S2 and S3 effectively constrain individual behaviors, leading to the observed reduction in inter-individual variation as most fish align with the stimulus.

In summary, our analysis reveals that no single individual consistently dominates the collective motion; instead, leadership is a transient and distributed property that shifts among members. However, the degree to which individuals respond to external cues can still vary. Investigating how these differences in responsiveness interact with the shifting leadership structure remains an important future direction for understanding the mechanisms of collective guidance.

\section{Conclusion}

In this paper, we presented an extended interaction model that incorporates external visual stimuli to estimate the dynamic interaction networks of schooling fish. By utilizing model-derived, interpretable indices at both the group and individual levels, we quantified behavioral dynamics and identified influential individuals within the collective. Our experiments with projector-based visual stimuli demonstrated that these indices effectively capture transient leadership shifts and the homogenization of behavioral responses under high-intensity stimuli. Specifically, while leadership is not fixed and shifts dynamically among members, strong external cues effectively constrain individual variability, leading to more uniform group alignment.

Because our model-derived indices can be computed in real time, they provide a powerful framework for the dynamic intervention and guidance of biological collectives. The ability to identify influential or stimulus-sensitive individuals in real time opens new possibilities for closed-loop swarm-machine interaction systems. Future work will focus on leveraging these insights to develop adaptive control strategies that can guide collective behavior by selectively targeting key individuals within the group.

\section*{Acknowledgment}

This work was supported by JSPS KAKENHI Grant Number JP21H05302.


\begin{thebibliography}{10}

\bibitem{bolstadCausalNetworkInference2011}
Andrew Bolstad, Barry Van~Veen, and Robert Nowak.
\newblock Causal network inference via group sparse regularization.
\newblock {\em IEEE Transactions on Signal Processing}, 59(6):2628--2641, June 2011.

\bibitem{egerstedtInteractingNetworksMobile2014a}
Magnus Egerstedt, Jean-Pierre {de~la Croix}, Hiroaki Kawashima, and Peter Kingston.
\newblock Interacting with networks of mobile agents.
\newblock In Peter Benner, Rolf Findeisen, Dietrich Flockerzi, Udo Reichl, and Kai Sundmacher, editors, {\em Large-{{Scale Networks}} in {{Engineering}} and {{Life Sciences}}}, pages 199--224. {Springer International Publishing}, {Cham}, 2014.

\bibitem{fujiiLearningInteractionRules2021}
Keisuke Fujii, Naoya Takeishi, Kazushi Tsutsui, Emyo Fujioka, Nozomi Nishiumi, Ryoya Tanaka, Mika Fukushiro, Kaoru Ide, Hiroyoshi Kohno, Ken Yoda, Susumu Takahashi, Shizuko Hiryu, and Yoshinobu Kawahara.
\newblock Learning interaction rules from multi-animal trajectories via augmented behavioral models.
\newblock {\em Advances in Neural Information Processing Systems}, 2021.

\bibitem{yolo11_ultralytics}
Glenn Jocher and Jing Qiu.
\newblock {Ultralytics YOLO11}.
\newblock \url{https://github.com/ultralytics/ultralytics}, 2024.
\newblock Version 11.0.0, AGPL-3.0 License.

\bibitem{Kawashima23}
Hiroaki Kawashima.
\newblock Estimating networks of interaction in fish schools with a minimal model.
\newblock {\em BiRD (PerCom2023 Workshop)}, 2023.

\bibitem{kawashimaManipulabilityLeaderFollower2014}
Hiroaki Kawashima and Magnus Egerstedt.
\newblock Manipulability of leader\textendash follower networks with the rigid-link approximation.
\newblock {\em Automatica}, 50(3):695--706, 2014.

\bibitem{RajBioRxiv2025}
Raj~Rajeshwar Malinda, Saeko Takizawa, Akiyuki Koyama, Takayuki Niizato, Hitoshi Habe, and Hiroaki Kawashima.
\newblock Speed-controlled visual stimuli modulate fish collective dynamics.
\newblock {\em bioRxiv, \url{https://doi.org/10.64898/2025.12.05.692523}}, 2025.

\bibitem{marcinkevicsInterpretableModelsGranger2021}
Ri{\v c}ards Marcinkevi{\v c}s and Julia~E. Vogt.
\newblock Interpretable models for {{Granger}} causality using self-explaining neural networks.
\newblock {\em International Conference on Learning Representations}, January 2021.

\bibitem{rahmaniControllabilityMultiAgentSystems2009}
Amirreza Rahmani, Meng Ji, Mehran Mesbahi, and Magnus Egerstedt.
\newblock Controllability of multi-agent systems from a graph-theoretic perspective.
\newblock {\em SIAM Journal on Control and Optimization}, 48(1):162--186, 2009.

\bibitem{Reynolds1987}
Craig~W. Reynolds.
\newblock Flocks, herds and schools: A distributed behavioral model.
\newblock {\em ACM SIGGRAPH Computer Graphics}, pages 25--34, 1987.

\bibitem{rosenthalRevealingHiddenNetworks2015}
Sara~Brin Rosenthal, Colin~R. Twomey, Andrew~T. Hartnett, Hai~Shan Wu, and Iain~D. Couzin.
\newblock Revealing the hidden networks of interaction in mobile animal groups allows prediction of complex behavioral contagion.
\newblock {\em Proceedings of the National Academy of Sciences}, 112(15):4690--4695, April 2015.

\end{thebibliography}
\end{document}